\documentclass[
reprint, aps, prx,nofootinbib,longbibliography,floatfix,nobalancelastpage, superscriptaddress]{revtex4-2}

\usepackage{graphicx}
\usepackage[utf8]{inputenc}
\usepackage{indentfirst}
\usepackage{physics}
\usepackage{braket}
\usepackage{float}
\usepackage{amsmath}
\usepackage{epstopdf}
\usepackage{footnote} 
\usepackage{CJK}
\usepackage{esint}
\usepackage{color}
\usepackage[T1]{fontenc}
\usepackage{subfigure}
\usepackage{amsfonts}
\usepackage{footmisc}
\usepackage{scrextend}
\usepackage{multirow}
\usepackage[hyperfootnotes=false]{hyperref}
\usepackage[acronym]{glossaries}
\usepackage[normalem]{ulem}
\usepackage{silence}
\usepackage{bm}

\newcommand{\bzero}{\boldsymbol{0}}

\newcommand{\hc}{\text{h.c.}}
\newcommand*\diff{\mathop{}\!\mathrm{d}}

\newcommand{\bxi}{\boldsymbol{\xi}}

\begin{document}


\title{Entangling remote superconducting qubits via transducer-generated multi-time-bin states}

\author{Jing Wu}
\affiliation{Fermi National Accelerator Laboratory, Batavia, IL 60510, USA}
\author{Changqing Wang}
\affiliation{Fermi National Accelerator Laboratory, Batavia, IL 60510, USA}
\author{Andrew Cameron}
\affiliation{Fermi National Accelerator Laboratory, Batavia, IL 60510, USA}
\author{Silvia Zorzetti}
\email{zorzetti@fnal.gov}
\affiliation{Fermi National Accelerator Laboratory, Batavia, IL 60510, USA}

\begin{abstract}

Recent studies have shown long-distance entanglement using NV centers, atoms, and quantum dots with single-photon time-bin encoding. We propose a method to entangle remote superconducting qubits via microwave-optical transduction using multi-time-bin states. By adapting conventional entanglement swapping techniques, fidelity improves from $0.75$ to $0.98$ in transduction systems, and $0.66$ to $0.89$ in noisy channels. The protocol mitigates thermal noise without relying on purification and offers a practical path toward scalable, heterogeneous quantum systems.


\end{abstract}

\maketitle

\textit{Introduction--}Distributing entanglement between remote qubits is a key goal in the development of quantum networks. 
The entanglement establishes a quantum connection between two remote quantum memories, enabling essential tasks such as quantum teleportation~\cite{QuantumTP, time-bin-TP,Thomas:24,Lago-Rivera2023} and quantum sensing~\cite{Zhang_2021,Kevin-sensing2021,Danilin_2024}. 
While quantum controls and computing using circuit quantum electrodynamics (cQED) have witnessed significant improvements in efficiency in recent years due to the nonlinear Josephson effect, transmitting microwave quantum states over long distances remains challenging due to high loss and noise at room temperature. 
A promising approach involves converting microwave photons to optical photons~\cite{fan2018,Holzgrafe2020,Han2020,Mirhosseini2020,Wang2022,Kurokawa2022,Stockill2022,Sekine2024,Shen2024, sahu2023entangling} and performing optical entanglement swapping based on single-photon or time-bin states~\cite{time-bin1999,deRiedmatten2005,Transmon-time-bin,Huang2022,Inagaki:13,PRXQuantum.1.020317,Zo2024}. The performance of entanglement swapping via cavity-based microwave-optical transduction, as well as methods to increase the fidelity of entangled qubits beyond entanglement purification~\cite{Bennett1996,Deutsch1996}, remains relatively underexplored. \\
In this work, we present an entanglement swapping protocol based on qubit–time-bin states and demonstrate that the conventional two-time-bin encoding is not optimal.
By extending the protocol to multi-time-bin encoding—where a superconducting qubit is coupled to an optical state spanning more than two time bins—we show that entanglement swapping performance can be improved without requiring additional complex quantum operations.
This approach is compatible with microwave-optical transducers operating at lower repetition rates, enabling the entanglement of two remote superconducting qubits with markedly reduced infidelity caused by thermal noise photons. 
Time-bin encoding is naturally compatible with a variety of quantum computing systems. 
While this work focuses on entangling remote superconducting qubits, the same multi-time-bin entanglement swapping protocol can benefit other quantum platforms—such as NV centers, trapped ions, and quantum dots~\cite{Transmon-time-bin,Lee_2019,DiamondSpin2019,ilves-demand_2020,atomic-memory,Krutyanskiy2023,Ruf-color-centers,Pan2022}—by mitigating infidelity due to noise photons, including thermal photons, Raman scattering photons, and detectors' dark counts.
  
\textit{Qubit-time-bin states--}By utilizing the sideband transition of a transmon qubit, we can generate an entangled qubit and a microwave time-bin state. 
Let $\ket{g}$, $\ket{e}$, and $\ket{f}$ denote the ground, first excited, and second excited states of the transmon qubit, respectively. 
Let $\ket{n}$ represent the Fock state of the microwave cavity.
Qubit control is achieved through microwave pulses with frequencies $\omega_q$ and $\omega_q+\alpha$ for transitions between the states $\ket{g}\leftrightarrow\ket{e}$ and $\ket{e}\leftrightarrow\ket{f}$.
Additionally, the drive at frequency $2\omega_q+\alpha-\omega_r$ activates the Jaynes-Cummings Hamiltonian between the microwave cavity and the transmon qubit, enabling a sideband transition between the states $\ket{g}\ket{n}\leftrightarrow\ket{f}\ket{n-1}$\cite{sideband1,sideband2}. The qubit-time-bin state can be generated as follows \cite{Microwave-Time-bin}:
\begin{align*}
    &(\ket{g}+\ket{e})\ket{0}\rightarrow \ket{f}\ket{0}+\ket{e}\ket{0}\rightarrow \ket{g}\ket{1}+\ket{e}\ket{0}\rightarrow...\\
    &\ket{g}\ket{n}+\ket{e}\ket{0}\rightarrow(\ket{g}\ket{n}_{t_1}+\ket{e}\ket{0}_{t_1})\ket{0}\rightarrow...\\
    &\ket{g}\ket{n0}_{t_1 t_2}+\ket{e}\ket{0n}_{t_1 t_2}.
\end{align*}
In the second line, the photons in the microwave cavity are released by a superconducting quantum interference device (SQUID) at time $t_1$. 
After the release, the second round of transitions generates the second time bin at $t_2$. The time-bin state $\ket{n0}_{t_1t_2}$ represents an early bin containing $n$ photons and a late bin containing $0$ photons. The final qubit-time-bin entangled state can be expressed as:
\begin{align}
    \ket{\psi}&=\frac{1}{\sqrt{2}}(\ket{g}_Q\ket{n0}_{t_1 t_2}+\ket{e}_Q\ket{0n}_{t_1 t_2}).
    \label{eq:transmon-time-bin-state}
\end{align}
By implementing the sideband scheme again, we are able to generate a qubit with $k$ time bins as:
\begin{align*}
    &(\ket{g}+\ket{e})\ket{0}\rightarrow  ...\ket{g}\ket{1}_{t_1}+\ket{e}\ket{0}_{t_1}\rightarrow...\\
    &\ket{g}\ket{101...}_{t_1t_2t_3...}+\ket{e}\ket{010...}_{t_1t_2t_3...}
\end{align*}
The resulting entangled state is expressed as:
\begin{align}
    \ket{\psi_k} =\frac{1}{\sqrt{2}}(\ket{g}_Q\ket{101...}_T+\ket{e}_Q\ket{010...}_T),
    \label{eq:transmon-k-time-bin}
\end{align}
where $T=t_1t_2...t_k$ denotes the continuous-variable quantum system at time bin $t_1,...,t_k$. 
The qubit-time-bin entangled state in Eq.~\eqref{eq:transmon-k-time-bin} can also be generated by transmon qubits as well as various other platforms, such as quantum dots \cite{Lee_2019}, NV centers \cite{DiamondSpin2019}, and trapped ions \cite{Krutyanskiy2023}.

\begin{figure}[t]
    \centering
    \includegraphics[width=0.98\linewidth]{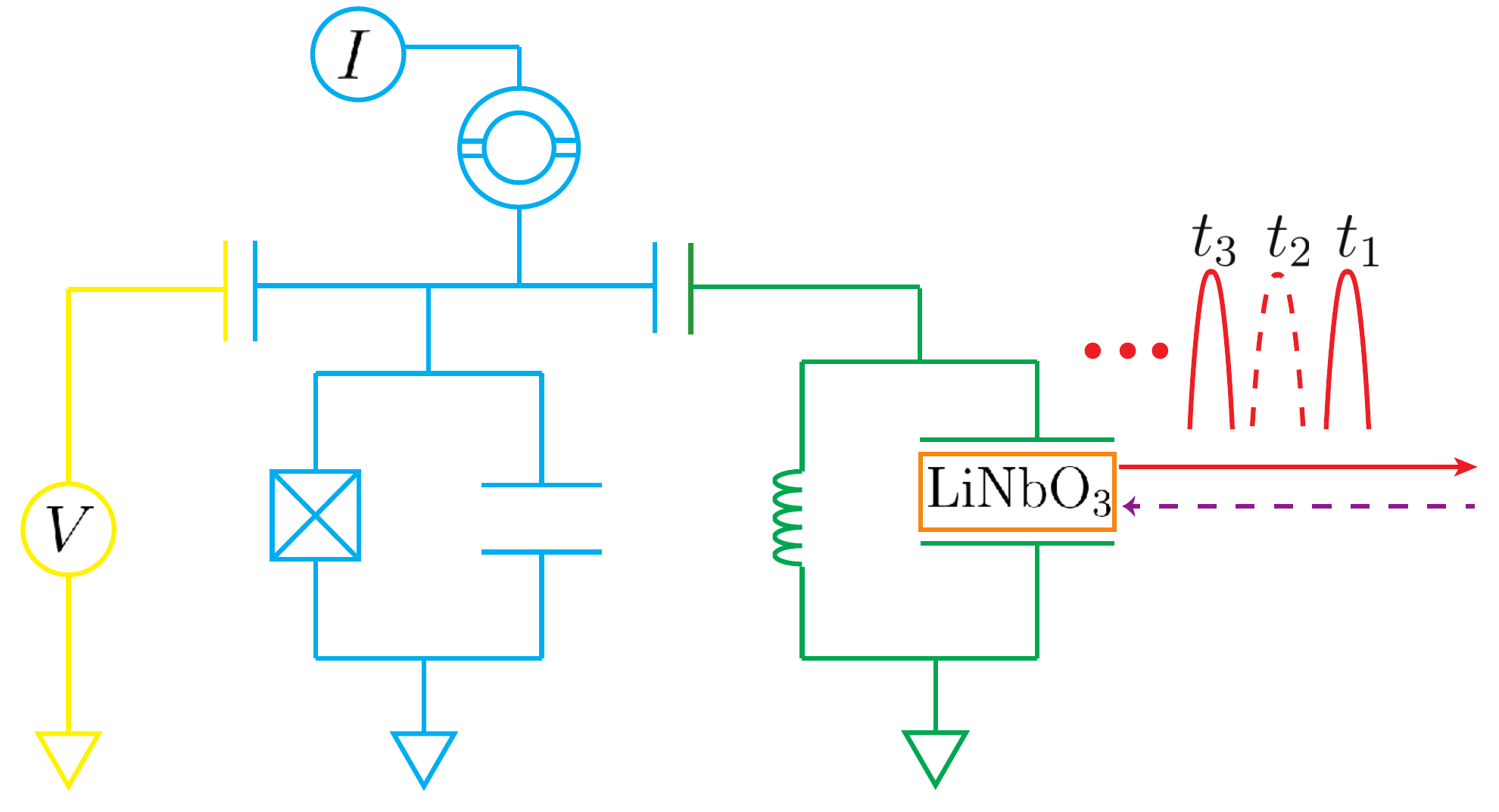}
    \caption{Generation of the transmon qubit-time-bin entangled state. $V$: Microwave drive for qubit control and sideband transition. $I$: SQUID for releasing microwave photons to the transducer.}
    \vspace{-3mm}
    \label{fig:sideband}
\end{figure}

\textit{State transfer and fidelity--}Converting microwave photons to optical photons can be achieved using a cavity-based microwave-optical quantum transducer operating at low temperatures and with a high quality factor, as shown in Fig.~\ref{fig:sideband}. 

The transduction system utilizes the electro-optic effect. In a typical direct-conversion approach, a beam-splitter-like interaction,
$
g \hat{m}^\dagger\hat{a} + \hc,
$
is induced between a microwave mode $\hat{m}$ and an optical mode $\hat{a}$ by a nonlinear crystal such as $\text{LiNbO}_3$. 
In cavity-based quantum transduction, the steady-state solution is obtained from the input-output relation of the transducer. The quantum channel is then derived as a Gaussian channel $\mathcal{N}_{\eta,N}$, with transmissivity $\eta$ and mixed-in noise $N = (1-\eta)(\bar{n} + 1/2)$, as given by~\cite{Tsang1,Tsang2},
\begin{subequations}
    \begin{align}
    &\eta = \zeta_{m} \zeta_{o} \frac{4C}{(1+C)^2}, \label{eq:eta_DC}\\
    &N=\frac{1}{2}+\frac{2C\zeta_o[2(1-\zeta_m)\Bar{n}_{\rm th}-\zeta_m]}{(1+C)^2}, \label{eq:N_DC}
    \end{align} 
    \label{eq:transducer-channel}
\end{subequations}
where $\zeta_{m} = \gamma_{mc}/(\gamma_{mc}+\gamma_{mi})$ and $\zeta_{o} = \gamma_{oc}/(\gamma_{oc}+\gamma_{oi})$ 
are the extraction efficiencies, with $\gamma_{mc}$ and $\gamma_{mi}$ denoting the microwave external loss rate and intrinsic loss rate, respectively. $C = 4g_{eo}^2 n_p/(\gamma_o \gamma_m)$ is the cooperativity, $\Bar{n}_{\rm th}$ is the mean photon number of microwave thermal noise, and $n_p$ is the pump photon number.
The Gaussian channel description can also be applied to other transduction systems or time-bin state transfer systems, where the interactions are linear and the noise state is thermal \cite{Rabl2010,Mirhosseini2020,Hatanaka2022,Sonar:25}.\\
 The performance of the transducer is evaluated by the quantum fidelity between the input pure state $\hat{\rho}^{\rm in}=\ketbra{\psi_k}$ and output mixed state $\hat{\rho}^{\rm out}=\mathcal{N}_{\eta,N}(\hat{\rho}^{\rm in})$. 
It represents the probability of obtaining the ideal state $\ket{\psi_k}$ from the noisy state $\hat{\rho}^{\rm out}$, and thus evaluates the rate of ideal entanglement swapping. 
The quantum fidelity is given by:
\begin{align}
    F&=\text{Tr}_{QT}(\hat{\rho}^{{\rm out}} \hat{\rho}^{{\rm in}}) \nonumber\\
    &= \frac{1}{\pi^k} \int \diff{\boldsymbol{\xi}}^{2k} \; \text{Tr}_{Q}\left[\rho^{{\rm out}}(\bxi) \rho^{{\rm in}}(-\bxi) \right],
    \label{eq:fidelity-general-expression}
\end{align}
where $\rho^{\rm out}(\bxi)$ and $\rho^{\rm in}(\bxi)$ are the $2\times2$ density matrices of the states $\hat{\rho}^{\rm out}$ and $\hat{\rho}^{\rm in}$, respectively. In the complex domain, the elements of the matrix are given by the quantum characteristic function:
\begin{align}
    \rho(\bxi) = \text{Tr}_T ({\hat{\rho}_{QT}}\hat{D}_T(\bxi )), \label{eq:characterstic-partial-trace}
\end{align}
where $\hat{D}_T(\bxi)$ is the displacement operator of the continuous-variable oscillators at time bins $T=t_1\dots t_k$. 
For a quantum transducer with 3D microwave cavity, designed mode coupling and low temperature, the external loss rates $\gamma_{oc}$ and $\gamma_{mc}$, can be up to $5$ to $10$ times higher than the intrinsic loss rates $\gamma_{mi}$ and $\gamma_{oi}$, resulting in large extraction efficiencies $\zeta_m$ and $\zeta_o$ \cite{fan2018, Wang2022, zorzetti2023millikelvin}. 
By operating at a lower repetition rate, the transducer can achieve higher cooperativity $C$ while maintaining the same level of thermal noise \cite{Rueda2019,Sahu2022}. Therefore, with an increased time spacing, $\Delta t=t_2-t_1$, synchronized with the reduced repetition rate, the quantum transducer enables the stable generation of a qubit-time-bin state. \\
The fidelity of the qubit-time-bin state for $k$ time bins is derived in Appendix~\ref{apd:quantum fidelity} and shown in Fig.~\ref{fig:time-bin-fidelity}.
Here, we assume a fixed noise photon number of $\bar{n}_{\rm th}=0.1$, which corresponds to the expected photon number at $180$~mK for $9$~GHz or $100$~mK for $5$~GHz, according to Bose–Einstein statistics.
The fidelity decreases as the number of time bins increases. However, as shown in the next section, entanglement swapping fidelity can improve with more time bins.

\begin{figure}[tp]
    \centering
   
\includegraphics[width=0.99\linewidth]{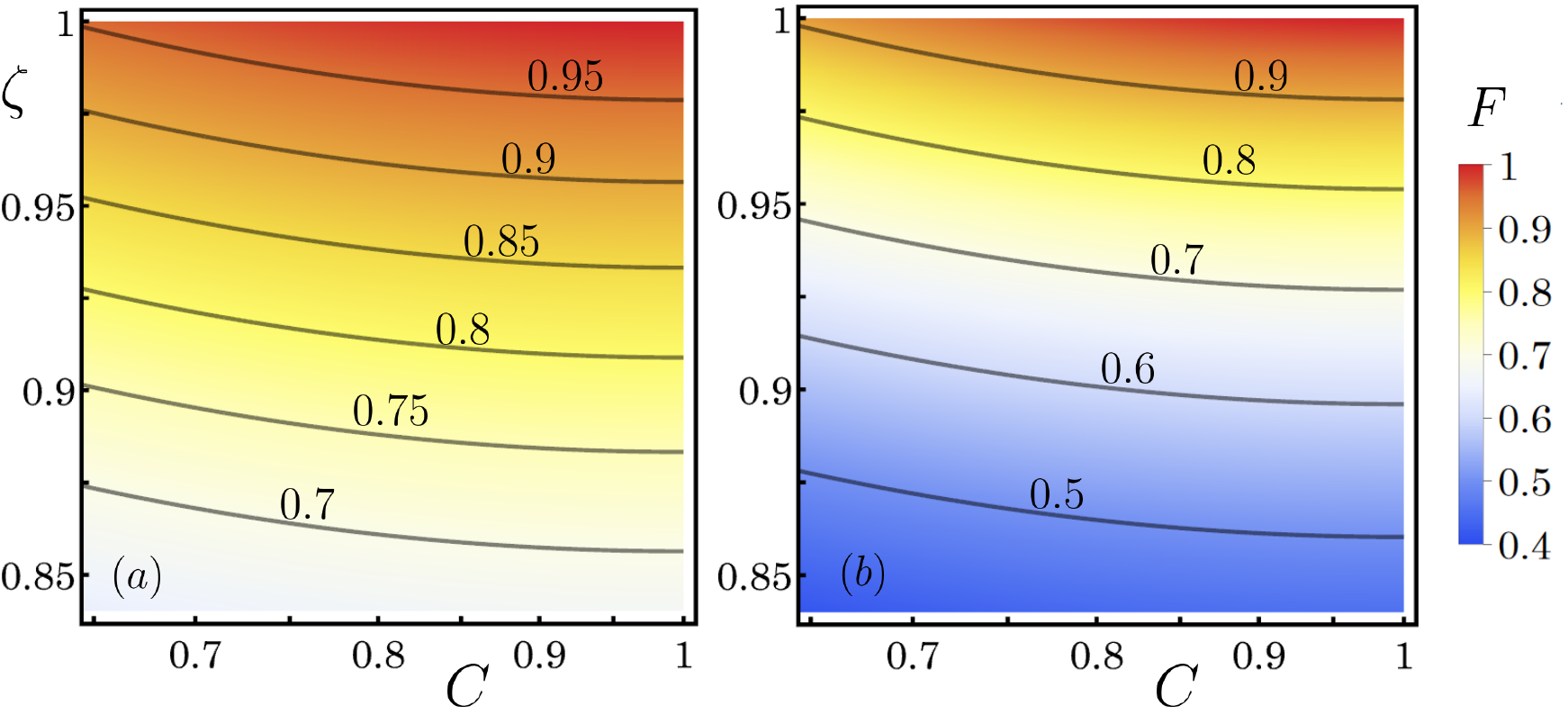}
    \caption{Fidelity of qubit-time-bin state versus cooperativity $C$ and extraction efficiency $\zeta_m=\zeta_o=\zeta$ when $\bar{n}_{\rm th}=0.1$: (a) two-time-bin state (b) four-time-bin state.}
    \vspace{-4mm}
    \label{fig:time-bin-fidelity}
\end{figure}

\textit{Entanglement swapping--}In this section, we illustrate entanglement swapping using Fock states and time-bin states. The process relies on a beam splitter and photon detection at each time bin. We evaluate the fidelity between the ideal qubit Bell state and the entangled qubits, which are heralded by the multi-time-bin Bell state.
From the results, we conclude that the fidelity can be reduced by increasing the number of time bins, thereby improving the quality of the entanglement.

\begin{figure}[tp]
    \centering
    \vspace{-0 mm}
    \includegraphics[width=0.95\linewidth]{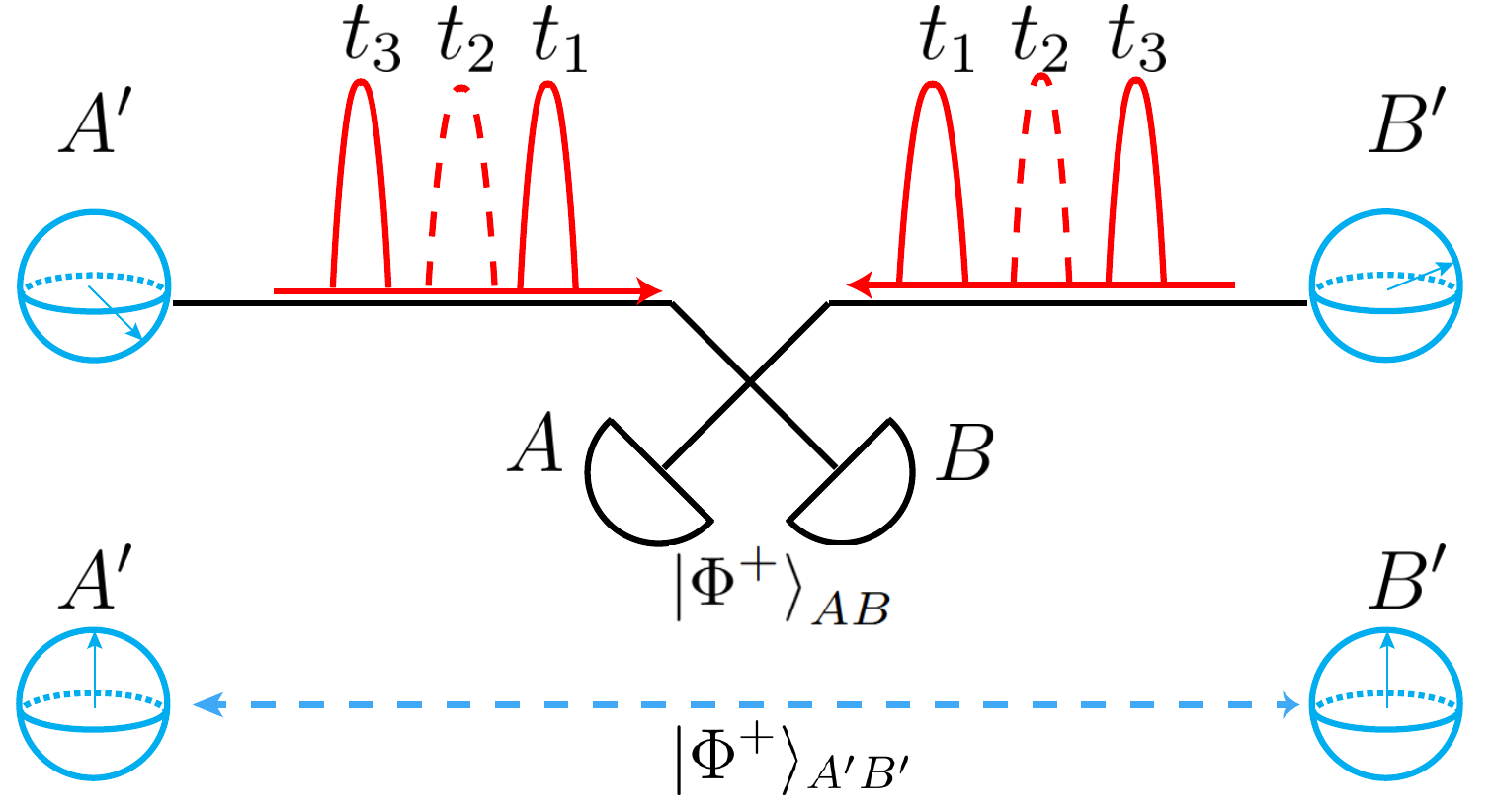}
    \caption{Entanglement swapping scheme: For two-time-bin swapping, the time-bin states are $\ket{n0}_{t_1t_2}$ and $\ket{0n}_{t_1t_2}$, while for three-time-bin swapping, the time-bin states are $\ket{101}$ and $\ket{010}$.}
    \vspace{-3mm}
    \label{fig:entanglement-swap}
\end{figure}

\begin{figure}[tpb]
    \centering
\includegraphics[width=\linewidth]{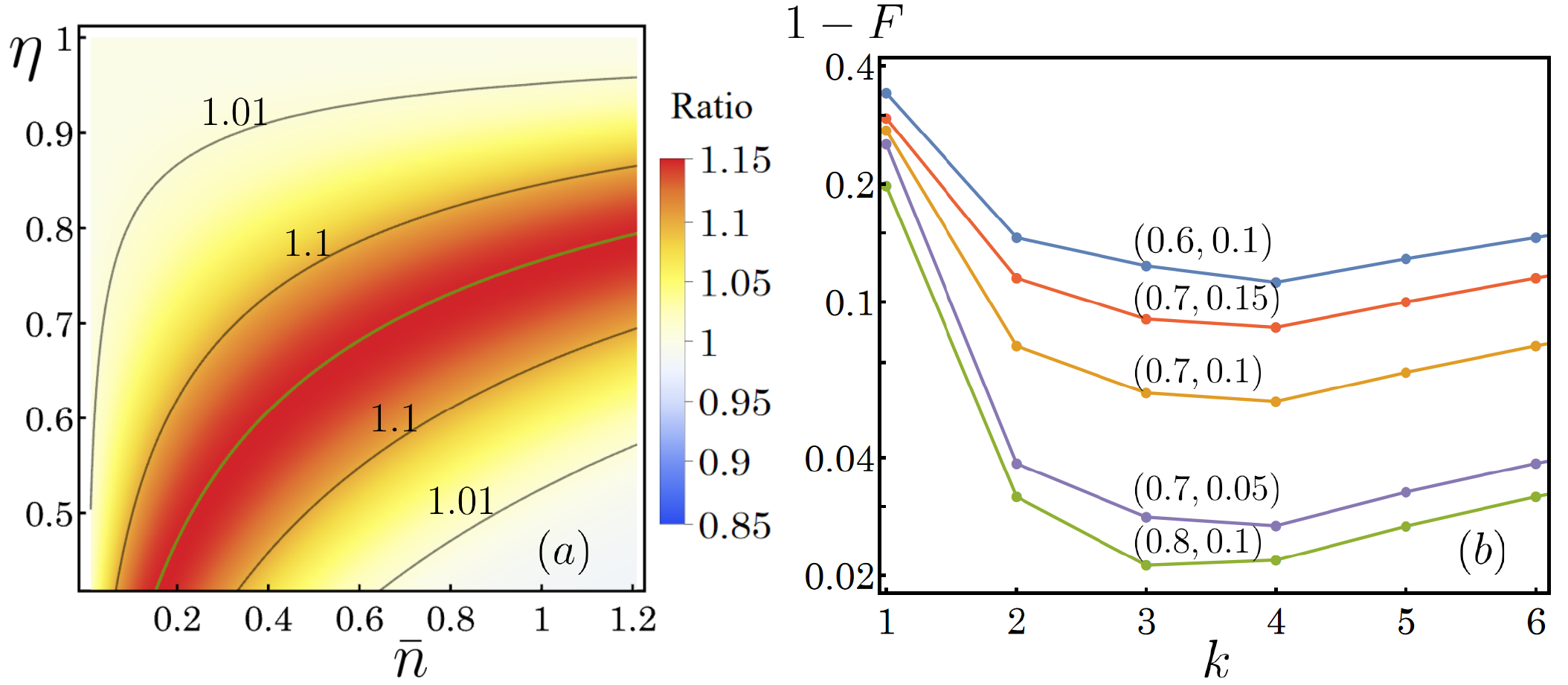}
    \caption{(a) Fidelity ratio  \(F(n=1)/F(n=2)\) as a function of \(\eta\) and \(\bar{n}\) for transduction Gaussian channels. The green curve corresponds to $F(n=2)=0.5$.
    (b) Infidelity as a function of the number of time bins $k$ for Gaussian channels. We use the notation $(\eta, \bar{n}) = (0.6, 0.1)$.}
    \vspace{-3mm}
    \label{fig:two-photon}
\end{figure}

The fundamental entanglement swapping protocol by Fock states and time-bin states is illustrated in Fig.~\ref{fig:entanglement-swap}.
Let $\ket{E}=\ket{n0}_{t_1t_2}$ and $\ket{L}=\ket{0n}_{t_1 t_2}$ be the early and late time-bin state encoded by Fock state $\ket{n}$, respectively. 
Alice prepares her state $1/\sqrt{2}( \ket{g}_{A^\prime}\ket{E}_{A}+\ket{e}_{A^\prime}\ket{L}_{A})$ and Bob prepares $ 1/\sqrt{2}(\ket{g}_{B^\prime}\ket{E}_{B}+\ket{e}_{B^\prime}\ket{L}_{B})$. 
The joint state of Alice and Bob can then be expressed in the Bell-state basis as follows:
\begin{align}
&1/2(\ket{g}_{A^\prime}\ket{E}_{A}+\ket{e}_{A^\prime}\ket{L}_{A})(\ket{g}_{B^\prime}\ket{E}_{B}+\ket{e}_{B^\prime}\ket{L}_{B})\nonumber\\
&=1/2\left[\ket{\Psi^+}_{A^\prime B^\prime}\ket{\Psi^+}_{AB}+\ket{\Psi^-}_{A^\prime B^\prime}\ket{\Psi^-}_{AB} \right. \nonumber\\
    &\qquad \qquad \left. +\ket{\Phi^+}_{A^\prime B^\prime}\ket{\Phi^+}_{AB}+\ket{\Phi^-}_{A^\prime B^\prime}\ket{\Phi^-}_{AB}\right],
    \label{eq:product-Bell-state}
\end{align}
where $\ket{\Psi^\pm}_{A^\prime B^\prime}$ and $\ket{\Phi^\pm}_{A^\prime B^\prime}$ are qubit Bell states, and  $\ket{\Psi^\pm}_{AB}$ and $\ket{\Phi^\pm}_{AB}$ are time-bin Bell states, given by
\begin{subequations}
\begin{align}
    &\ket{\Psi^\pm}_{A^\prime B^\prime}=1/\sqrt{2}(\ket{gg}\pm\ket{ee}),\\
    &\ket{\Phi^\pm}_{A^\prime B^\prime}=1/\sqrt{2}(\ket{ge}\pm\ket{eg}),\\    
    &\ket{\Psi^\pm}_{AB}=1/\sqrt{2}(\ket{EE}\pm\ket{LL}),\\
    &\ket{\Phi^\pm}_{AB}=1/\sqrt{2}(\ket{EL}\pm\ket{LE}).
    \label{eq:time-bin-bell}
\end{align}
\end{subequations}
Photon number detections at the two output ports of the beam splitter $AB$ at time bins $t_1$ and $t_2$ herald the time-bin Bell states $\ket{\Phi^+}_{AB}$ and $\ket{\Phi^-}_{AB}$, which are encoded by the Fock state.
In general, all detection events can be grouped into three classes, as illustrated in Appendix~\ref{apd:multi-photon swap}. 
Two of these classes correspond to orthogonal events that allow us to distinguish the time-bin Bell states $\ket{\Phi^+}_{AB}$ and $\ket{\Phi^-}_{AB}$, while the remaining class corresponds to $\ket{\Psi^\pm}_{AB}$, making these Bell states indistinguishable in this scheme.
Thus, by Eq.~\eqref{eq:product-Bell-state}, the detection of $\ket{\Phi^\pm}_{A B}$ heralds the qubit Bell state $\ket{\Phi^\pm}_{A^\prime B^\prime}$ between Alice and Bob.\\
The entanglement swapping with a non-ideal state, given by $\hat{\rho}^{\rm out}=\mathcal{N}_{\eta,N}(\hat{\rho}^{\rm in})$, is illustrated in Appendix \ref{apd:multi-photon swap} for $\ket{n=1}$ and $\ket{n=2}$ Fock states. 
Let $\ket{\boldsymbol{n}}$ represent the photon number detection event, such as $\ket{2020}_{A_1 B_1 A_2 B_2}$. The $4\times4$ density matrix of the qubit state $\rho_{A^\prime B^\prime}$ is obtained as follows:
\begin{align}
     \rho_{A^\prime B^\prime}\propto \int \diff{\bxi}^{4k} \; \chi_{\hat{M}}(-\bxi) (\rho^{\rm out}_{A A^\prime}\otimes \rho^{\rm out}_{B B^\prime}),
     \label{eq:rho-AB}
\end{align}
where $\hat{M}=U_{\rm BS}^\dagger\ketbra{\boldsymbol{n}}U_{\rm BS}$, $U_{\rm BS}$ is the Gaussian unitary of the beam splitter, and $k$ is the number of time bins. 
For a Gaussian channel with transmissivity $\eta$ and environment noise $\Bar{n}=N/(1-\eta)+1/2$, the fidelity $F=\bra{\Phi^+}\hat{\rho}_{A^\prime B^\prime}\ket{\Phi^+}_{A^\prime B^\prime}$ is obtained for Fock states $\ket{n=1}$ and $\ket{n=2}$, as given by Eqs.~\eqref{eq:fidelity-single-photon} and \eqref{eq:fidelity-two-photon}, respectively.\\
The ratio $F(n=1)/F(n=2)$ as a function of $\eta$ and $\Bar{n}$ is plotted in Fig. \ref{fig:two-photon}(a). 
The region above the green curve corresponds to a fidelity $F(n=2)>0.5$. 
In this practical regime—where evident entanglement can be generated—we observe that the fidelity of entangled qubits obtained via single-photon time-bin state swapping exceeds that of the two-photon state.
As the photon number $n$ increases, the qubit fidelity decreases, since the coherence of the Fock state $\ket{n}$ is preserved with a probability that scales as $\eta^k$.

Therefore, the entanglement swapping using the single-photon state ($\ket{1}$) outperforms schemes based on higher Fock states ($\ket{n > 1}$) in each time bin.
Accordingly, this section focuses on single-photon states distributed across multiple time bins.
The early and late states in Eq. (\ref{eq:product-Bell-state}) are now $\ket{10...}_{t_1t_2...}$ and $\ket{01...}_{t_1t_2...}$, respectively.
The case $k=1$ corresponds to entanglement swapping with a single photon and the Hong–Ou–Mandel dip, while $k=2$ corresponds to entanglement swapping using early and late time bins.
From Eq.~\eqref{eq:time-bin-bell}, the $k$-time-bin Bell state is given by
\begin{align}
    \ket{\Phi^\pm}_{AB} =1/\sqrt{2}(& \ket{10}_{A_1B_1}\ket{01}_{A_2B_2}...\nonumber\\
    &\pm\ket{01}_{A_1B_1}\ket{10}_{A_2B_2}...),
    \label{eq:Bell-state1}
\end{align}
where $A_i$ is the $i$-th time bin at port $A$. In the ideal entanglement swapping scheme, the $50/50$ beam splitter transforms the annihilation operators of the $i$-th time bin as follows:
\begin{align*}
    &\hat{a}_{A_i} \rightarrow 1/\sqrt{2} (\hat{a}_{A_i} + \hat{a}_{B_i}),\\
    &\hat{a}_{B_i} \rightarrow 1/\sqrt{2}(\hat{a}_{A_i} - \hat{a}_{B_i}).
\end{align*}
After the beam splitter, the Bell state in Eq.~\eqref{eq:Bell-state1} becomes 
\begin{align}
    U_{\rm BS}\ket{\Phi^\pm}_{AB}&=\frac{1}{(\sqrt{2})^{k+1}}\left[\underbrace{(\hat{a}_{A_1}^\dagger+\hat{a}_{B_1}^\dagger)(\hat{a}_{A_2}^\dagger-\hat{a}_{B_2}^\dagger)...}_{\substack{P_1}} \right.\nonumber\\
    &\left. \quad \pm\underbrace{(\hat{a}_{A_1}^\dagger-\hat{a}_{B_1}^\dagger)(\hat{a}_{A_2}^\dagger+\hat{a}_{B_2}^\dagger)...}_{\substack{P_2}}\right]\ket{\boldsymbol{0}}.
\end{align}
Performing single photon detections at all $A_iB_i$, there are $2^k$ orthogonal events corresponding to $U_{\rm BS}\ket{\Phi^\pm}_{AB}$.
Each event has an equal probability of detection and they can be classified into two subgroups by parities $P_1$ and $P_2$. 
Initialize $P_1=P_2=1$.
At the $i$-th time bin, the joint state $A_iB_i$ must be either $\ket{01}$ or $\ket{10}$. 
If $\ket{01}$ is detected at an odd-numbered time bin, update $P_2\rightarrow-P_2$. If $\ket{01}$ is detected at an even-numbered time bin, update $P_1\rightarrow-P_1$.
If $P_1=P_2$ at the end, the detection events correspond to the Bell state $\ket{\Phi^+}_{AB}$; otherwise, they correspond to $\ket{\Phi^-}_{AB}$.
\\
The fidelity of the entangled qubits, given a non-ideal time-bin state transmitted through a Gaussian channel, is evaluated based on the detection of the event $\ket{\boldsymbol{n}}=\ket{1010...}_{A_1B_1...}$. We observe that this event corresponds to the Bell state $\ket{\Phi^+}_{AB}$. 
Let $\hat{M}=U_{\rm BS}^\dagger\ketbra{\boldsymbol{n}} U_{\rm BS}$.
By Eq.~\eqref{eq:characterstic-partial-trace}, we obtain the characteristic function of $AA^\prime$ and $\hat{M}$:
\begin{align}
    &\rho^{\rm out}_{A A^\prime} = \frac{1}{2}e^{-(\eta/2+N)\sum_i |\xi_{A_i}|^2} \nonumber\\
    &\resizebox{0.42\textwidth}{!}{%
    $\begin{pmatrix}
        \Pi_{i \text{ odd}} L_1(\eta|\xi_{A_i}|^2) & \eta^{\frac{k}{2}} (\Pi_{i \text{ odd}}-\xi_{A_i}^*)(\Pi_{i \text{ even}}\xi_{A_i}) \\
        \eta^{\frac{k}{2}} (\Pi_{i \text{ odd}} \xi_{A_i})(\Pi_{i \text{ even}}-\xi_{A_i}^*)  & \Pi_{i \text{ even}} L_1(\eta|\xi_{A_i}|^2)
    \end{pmatrix}$,
    } \nonumber \\
    &\chi_{\hat{M}}  (\bxi)=e^{-1/2(\sum_i|\xi_{A_i}|^2+|\xi_{B_i}|^2)}\;\Pi_i L_1(1/2|\xi_{A_i}-\xi_{B_i}|^2), \nonumber
\end{align}
where $L_n$ is the n-th Laguerre polynomial. Substituting them into Eq.~\eqref{eq:rho-AB}, the fidelity of the entangled qubit can be obtained from
\begin{subequations}
\begin{align}
    &K_0=[\rho_{A^\prime B^\prime}]_{11}+[\rho_{A^\prime B^\prime}]_{22}+[\rho_{A^\prime B^\prime}]_{33}+[\rho_{A^\prime B^\prime}]_{44}, \label{eq:K0} \\
    &K_0 F
    =1/2([\rho_{A^\prime B^\prime}]_{22}+[\rho_{A^\prime B^\prime}]_{33}+[\rho_{A^\prime B^\prime}]_{23}+[\rho_{A^\prime B^\prime}]_{32}),
    \label{eq:K0F}
\end{align}
\end{subequations}
where $[\rho_{A^\prime B^\prime}]_{ij}$ denotes the element of $\rho_{A^\prime B^\prime}$ at $i$-th row and $j$-th column. The normalization constant and the fidelity are listed in the Appendix~\ref{apd:fidelity-multiple-time-bin}.

\begin{figure}[tp]
    \centering
\includegraphics[width=\linewidth]{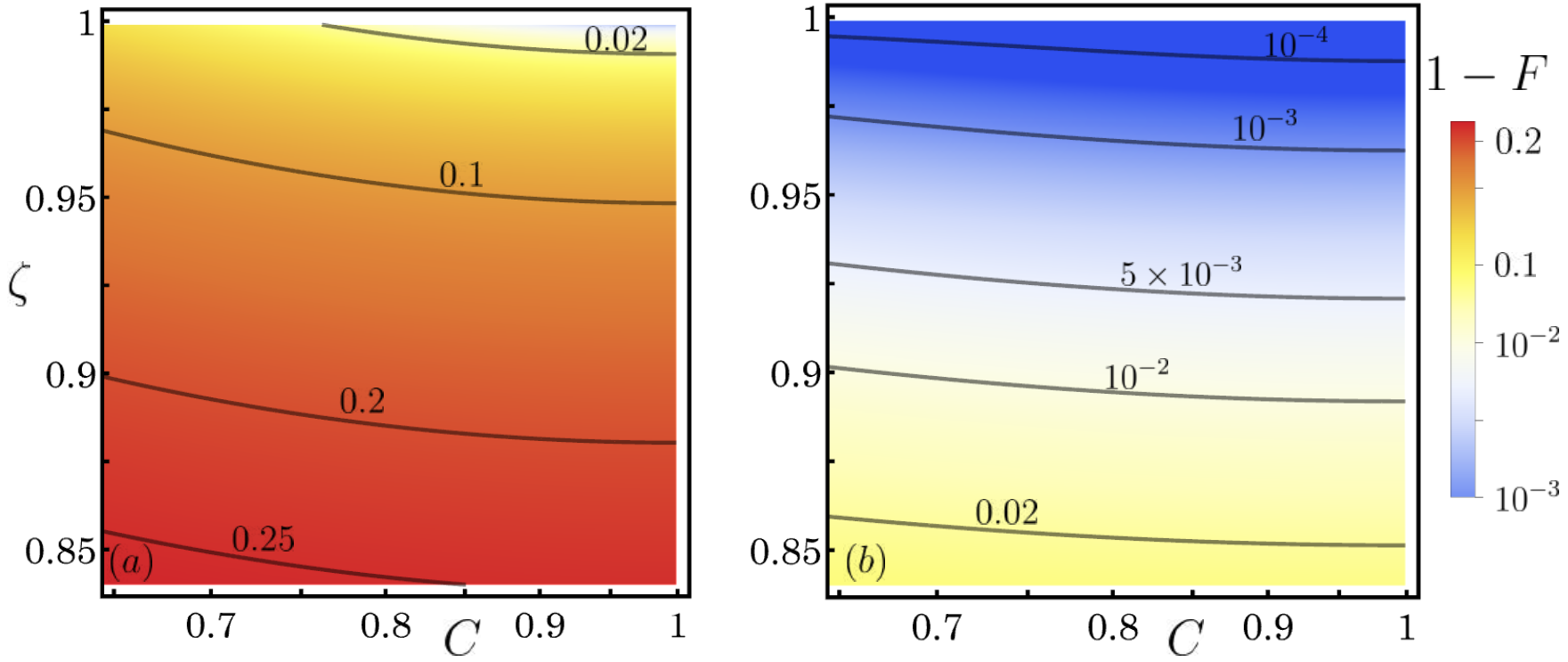}
    \caption{Infidelity versus the extraction efficiency $\zeta$ and cooperativity $C$ for (a) a single time bin, and (b) $k$ time bins, where $k$ is chosen to minimize the infidelity.
    }
    \label{fig:fidelity-time-bin}
    \vspace{-3 mm}
\end{figure}

From these calculations, we have
\begin{align}
    & [\rho_{A^\prime B^\prime}]_{11}\propto\left\{ \frac{(t-1)(t - \eta) \left[ t^2 + 2\eta - t (1 + \eta) \right]}{t^8} \right\}^{k/2},\label{eq:rho11}\\
    &\frac{[\rho_{A^\prime B^\prime}]_{23}}{[\rho_{A^\prime B^\prime}]_{33}}=\frac{1}{2}  +\frac{1}{2} \left[ \frac{\eta}{2 t ( t - \eta-1) + 3 \eta} \right]^k, \label{eq:rho23-33}
\end{align}
where $t=1+(1-\eta)\Bar{n}\geq1$. 
For pure loss channels $\Bar{n}=0$, $[\rho_{A^\prime B^\prime}]_{11}=[\rho_{A^\prime B^\prime}]_{44}=0$ and $[\rho_{A^\prime B^\prime}]_{23}=[\rho_{A^\prime B^\prime}]_{33}$, leading to the ideal heralded state $\ket{\Phi^+}_{A^\prime B^\prime}$.
In contrast, for a thermal-loss channel $\Bar{n}>0$, the noise photons introduce both depolarization as in Eq.~\eqref{eq:rho11} and decoherence as in Eq.~\eqref{eq:rho23-33}.
Increasing the number of time bins effectively suppresses the terms $[\rho_{A^\prime B^\prime}]_{11}$ and $[\rho_{A^\prime B^\prime}]_{44}$, much faster than the ratio $[\rho_{A^\prime B^\prime}]_{23}/[\rho_{A^\prime B^\prime}]_{33}$. 
Thus, depolarization is suppressed rapidly, while decoherence changes gradually. An optimal number of time bins, between $1$ and infinity, can therefore be selected to minimize the infidelity of the quantum state.\\
The infidelity of the entangled qubits as a function of the number of time bins is shown in Fig.~\ref{fig:two-photon}(b) for Gaussian channels. For a noisy Gaussian channel with $\eta=0.6$ and $\Bar{n}=0.1$, the fidelity increases from $0.66$ to $0.89$ with four time bins. And for $\eta=0.8$ and $\Bar{n}=0.1$, the fidelity increases from $0.8$ to $0.98$ with three time bins.\\
The multi-time-bin approach can be applied to entangle superconducting qubits via microwave-optical transducers, which introduce thermal-loss channels described by Eqs.~\eqref{eq:transducer-channel}. 
We take thermal photon number $\bar{n}_{\rm th}=0.1$. 
The infidelity of the resulting entangled state is evaluated as a function of the extraction efficiency $\zeta$ and cooperativity $C$, as shown in Fig.~\ref{fig:fidelity-time-bin}.
From the figure, infidelity decreases with higher extraction efficiency and cooperativity. Comparing Figs.~\ref{fig:fidelity-time-bin}(a) and \ref{fig:fidelity-time-bin}(b), multi-time-bin encoding effectively suppresses infidelity, reducing it from $0.25$ to $0.02$ and from $0.1$ to $5\times10^{-3}$.
In the parameter regime of 
$\zeta$ and 
$C$ considered in our simulations, increasing the number of time bins is more efficient than entanglement purification protocols like BBPSSW and DEJMPS, which typically require $4$ to $6$ iterations and an exponentially growing number of entangled state copies to increase the fidelity from $0.75$ to $0.98$.

\textit{Discussion--}We present a theoretical study of entanglement swapping using time-bin states encoded in Fock states. 
Our findings show that encoding multiple time bins in single photons can enhance the fidelity of the entangled qubit after swapping, in the presence of thermal noise photons. The fidelity as a function of the number of time-bin modes is not a monotonic function. It is optimized for a fixed number of time-bin modes, after which, the infidelity slightly increases. 

This improvement occurs because the depolarization terms $[\rho_{A^\prime B^\prime}]_{11}$ and $[\rho_{A^\prime B^\prime}]_{44}$ decay much faster than the decoherence term $[\rho_{A^\prime B^\prime}]_{23}/[\rho_{A^\prime B^\prime}]_{33}$ as the number of time bins increases. 
Meanwhile, the entanglement swapping rate, as evaluated by the fidelity of the source, decreases moderately with the number of time bins. Although the rate of decrease may not be the primary limitation, the improvement in fidelity is ultimately bounded by Eq.~\eqref{eq:rho23-33}. \\
This approach is applicable to entangling different types of qubits. Here, we explore its application to cavity-based microwave-optical transducers, where thermal photons are inevitably introduced by the optical pump. 
By using multiple time bins instead of a single time bin, the infidelity can typically be reduced from 
$0.2$ to $10^{-2}$ when the microwave-optical quantum transducer is operating with
$\zeta=0.9$, $C=0.65$, and within the temperature range of $100-200$~mK. Although we demonstrate that higher-number Fock states do not outperform single-photon states, further investigation is needed to determine whether other engineered states and detection schemes could improve the rate and fidelity of entanglement swapping.


\textit{Acknowledgments--}This work is supported by the Fermi Forward Discovery Group LLC under Contract No. FWP-23-24 with the U.S. Department of Energy, Office of Science, Advanced Scientific Computing Research (ASCR) Program. S.Z. and C.W. also acknowledge support by the U.S. Department of Energy, Office of Science, National Quantum Information Science Research Centers, Superconducting Quantum Materials and Systems Center (SQMS) under Contract No. DE-AC02-07CH11359, and the DOE's Early Career Research Program. The NQI Research Center SQMS contributed by providing access to the experimental facilities. 


\onecolumngrid
\bibliography{ref.bib}

\clearpage

\setcounter{page}{1}
\appendix
\twocolumngrid

\section{Fidelity of qubit-time-bin state}
\label{apd:quantum fidelity}
Let the input state be $\hat{\rho}^{\rm in} = \ketbra{\psi_k}$, with $\ket{\psi_k}$ as given in Eq.~\eqref{eq:transmon-k-time-bin}. For the time-bin $t_i$, the characteristic function is defined as $\chi_{\hat{\rho}} = \Tr
_{t_i}(\hat{\rho} \hat{D}(\xi_i))$, where $\xi_i$ corresponds to the $i$-th time bin of the $k$ time bins. For $k$ time bins, we have $k$ complex variables, and the input state density matrix is given by $\rho^{\rm in} = \text{Tr}_{t_1 \dots t_k}(\hat{\rho} \hat{D}_T(\bxi))$. The density matrix of the input state is obtained as
\begin{align}
    \rho^{\rm in} = 1/2\begin{pmatrix}
        \chi_{\ketbra{1}}(\xi_1) ... & \chi_{\ketbra{1}{0}} (\xi_1) ...\\
        \chi_{\ketbra{0}{1}} (\xi_1) ... & 
        \chi_{\ketbra{0}}(\xi_1) ...
    \end{pmatrix},\nonumber
\end{align}
where the characteristic functions in the complex domain are given by
\begin{subequations}
\begin{align}
    &\chi_{\ketbra{n}} (\xi) = \Tr_T{[\ketbra{n}\hat{D}(\xi)]}=L_n(|\xi|^2)e^{-|\xi|^2/2},\\
    &\chi_{\ketbra{0}{n}} (\xi) =\Tr_T{[\ketbra{0}{n}\hat{D}(\xi)]} = \frac{\xi^n}{\sqrt{n!}}e^{-|\xi|^2/2},\\
    & \chi_{\ketbra{n}{0}} (\xi) = \chi_{\ketbra{0}{n}}^* (-\xi),
\end{align}
\label{eq:charactersitc-fock}
\end{subequations}
where $L_n$ is the n-th Laguerre polynomial.
For a Gaussian channel $\hat{\rho}^{{\rm out}}=\mathcal{N}_{\eta,N}(\hat{\rho}^{\rm in })$, we use the following properties of characteristic function
\begin{subequations}
\begin{align}
    &\chi^{\text{out}}(\xi) = \chi^{\text{in}}(\sqrt{\eta} \, \xi) \, e^{- N |\xi|^2},
    \label{eq:characteristic-input-output}\\
    &    {\text{Tr}} (\hat{\sigma}\hat{\rho})= \frac{1}{\pi^k} \int\diff^{2k}\boldsymbol{\xi}\; \chi_{\hat{\rho}}(\boldsymbol{\xi})\chi_{\hat{\sigma}}(\boldsymbol{-\xi}).
    \label{eq:charactersitc-fidelity}
\end{align}
\end{subequations}
to find the density matrix of the output state $\rho^{\rm out}$ and the quantum fidelity between the input and output state. 
Substituting the characteristic functions in Eqs.~\eqref{eq:charactersitc-fock} with $n=1$ into Eq. (\ref{eq:characteristic-input-output}), we obtain the density matrix
\begin{align}
    &\rho^{\rm out} = \frac{1}{2}e^{-(\eta/2+N)\sum_i |\xi_i|^2} \nonumber\\
    &\resizebox{0.42\textwidth}{!}{%
    $\begin{pmatrix}
        \Pi_{i \text{ odd}} L_1(\eta|\xi_i|^2) & \eta^{\frac{k}{2}} (\Pi_{i \text{ odd}}-\xi_i^*)(\Pi_{i \text{ even}}\xi_i) \\
        \eta^{\frac{k}{2}} (\Pi_{i \text{ odd}} \xi_i)(\Pi_{i \text{ even}}-\xi_i^*)  & \Pi_{i \text{ even}} L_1(\eta|\xi_i|^2)
    \end{pmatrix}$.
    \label{eq:rho-out-time-bin}
    }
\end{align}
The density matrix of the input state corresponds to a special case when $\eta = 1$ and $N = 0$.
The quantum fidelity is then obtained by taking the trace of the product of the density matrices $\rho^{\rm in}\rho^{\rm out}$, which involves integrating the characteristic functions as shown in Eq.~\eqref{eq:fidelity-general-expression}.
When $k=2l+1$ is odd, then
\begin{align}
    F=\left[\frac{t^2+2\eta-t(1+\eta)}{t^4}\right]^l\frac{2t^2+2\eta-t(1+\eta)}{4t^3}+\frac{\eta^{\frac{k}{2}}}{2t^{2k}},
\end{align}
where $t=(1+\eta)/2+N$.
And when $k$ is even,
\begin{align}
        F=\frac{1}{2}\left[\frac{t^2+2\eta-t(1+\eta)}{t^4}\right]^{k/2}+\frac{\eta^{\frac{k}{2}}}{2t^{2k}}.
\end{align}
By the uncertainty principle of environment state, $\boldsymbol{N}\pm\frac{i}{2}(\boldsymbol{\Omega}-\boldsymbol{T}\boldsymbol{\Omega}\boldsymbol{T}^T)\geq 0$, we obtain $N\geq (1-\eta)/2$ and $t\geq1$ with equality when the channel is pure loss.
Thus in general, the fidelity decreases when $\eta $ decreases or $k$ increases.

\section{Swapping by Fock states}
\label{apd:multi-photon swap}
\begin{table}[tb]
\caption{Detection events of single-photon swapping}\label{table:single-swap}
\centering
\begin{tabular}{ |c|c|c|c|c| } 
\hline
Product state & $A_1$ (early) & $B_1$ (early) & $A_2$ (late)&  $B_2$(late)\\
\hline
\multirow{3}{4em}{$\ket{\Phi^+_{AB}}$} & 1 & 0 &1 &0 \\ 
& 0 & 1 &0 &1 \\ 
\hline
\multirow{3}{4em}{$\ket{\Phi^-_{AB}}$} &0 & 1 &1 &0 \\ 
&1 & 0 &0 &1 \\
\hline
\multirow{3}{4em}{$\ket{\Psi^\pm_{AB}}$} &2 & 0 &0 &0 \\ 
 &0 & 2 &0 &0 \\
  &0 & 0 &2 &0 \\ 
   &0 & 0 &0 &2 \\ 
\hline
\end{tabular}
\end{table}

\begin{table}[tb]
\caption{Detection events of two-photon swapping}\label{table:two-swap}
\centering
\begin{tabular}{ |c|c|c|c|c| } 
\hline
Product state &  $A_1$ (early) & $B_1$ (early) &$A_2$ (late) &$B_2$ (late)\\
\hline
\multirow{3}{4em}{$\ket{\Phi^+_{AB}}$} & 2 & 0 &2 &0 \\ 
& 0 & 2 &0 &2 \\ 
& 0 & 2 &2 &0 \\
& 2 & 0 &0 &2 \\
& 1 & 1 &1 &1 \\
\hline
\multirow{3}{4em}{$\ket{\Phi^-_{AB}}$} &2 & 0 &1 &1 \\ 
&0 & 2 &1 &1 \\
&1 & 1 &2 &0 \\
&1 & 1 &0 &2 \\
\hline
\multirow{3}{4em}{$\ket{\Psi^\pm_{AB}}$} &4 & 0 &0 &0 \\ 
&0 & 4 &0 &0 \\
&2 & 2 &0 &0 \\
&0 & 0 &4 &0 \\
&0 & 0 &0 &4 \\
&0 & 0 &2 &2 \\
\hline
\end{tabular}
\end{table}
As shown in Fig. \ref{fig:entanglement-swap}, to measure the time-bin Bell states, we first use a $50/50$ beam splitter, which transforms the annihilation operators as follows:
\begin{subequations}
\begin{align}
    &\hat{a}_{A_1} \rightarrow 1/\sqrt{2} (\hat{a}_{A_1} + \hat{a}_{B_1}),\\
    &\hat{a}_{B_1} \rightarrow 1/\sqrt{2}(\hat{a}_{A_1} - \hat{a}_{B_1}), 
\end{align}
\label{eq:BS-transform}
\end{subequations}
where we use the notation $A_1$ to denote the early bin at port $A$, and $B_1$ to denote the early bin at port $B$.
Consider the case where both early and late bins contain single photons. After the beam splitter transformation,
the time-bin Bell state 
\begin{align*}
\ket{\Phi^\pm_{AB}}=(\hat{a}^\dagger_{A_1}\hat{a}^\dagger_{B_2} \pm \hat{a}_{A_2}^\dagger\hat{a}_{B_1}^\dagger) \ket{\bzero},
\end{align*} 
transforms to
\begin{align}
    \hat{U}_{BS}\ket{\Phi^\pm_{AB}}&= [(\hat{X}^\dagger+\hat{Y}^\dagger)\pm(\hat{X}^\dagger-\hat{Y}^\dagger)] \ket{\bzero},
 \label{eq:Phi_CD}\\
\hat{X}&=\hat{a}_{A_1}\hat{a}_{A_2}-\hat{a}_{B_1}\hat{a}_{B_2},\nonumber\\
\hat{Y}&=\hat{a}_{B_1}\hat{a}_{A_2}-\hat{a}_{A_1}\hat{a}_{B_2}.\nonumber
\end{align}
If the photon detection events correspond to $\hat{X}$, we obtain the product state in Eq.~(\ref{eq:product-Bell-state}) in $\ket{\Phi^+_{AB}}$, thereby heralding the state $\ket{\Phi^+_{A^\prime B^\prime}}$. 
Similarly, if the photon detection events correspond to $\hat{Y}$, the state $\ket{\Phi^-_{A^\prime B^\prime}}$ is heralded.
We summarize detection events in Table~\ref{table:single-swap}.
When the early and late time-bin are encoded by Fock state $\ket{n}$, i.e.  $\ket{n0}$ and $\ket{0n}$, Eq.~(\ref{eq:Phi_CD})
becomes
 \begin{align*}
     \ket{\Phi^\pm_{AB}}&=[(\hat{X}^\dagger+\hat{Y}^\dagger)^n\pm(\hat{X}^\dagger-\hat{Y}^\dagger)^n] \ket{\bzero}.
 \end{align*}
Since $\ket{\Phi^+_{AB}}$ contains terms of the form $(\hat{X}^{\dagger})^k (\hat{Y}^{\dagger})^{n-k}$ for $k=n,n-2,...$, while $\ket{\Phi^-_{AB}}$ contains terms for $k=n-1,n-3,...$, the $\ket{\Phi^+_{AB}}$ and $\ket{\Phi^-_{AB}}$ can still be distinguished by photon number detections at port $A$ and $B$.
The detection events for the two-photon swapping, $n=2$, are summarized in Table~\ref{table:two-swap}. Furthermore, we are able to distinguish $\ket{\Phi^\pm_{\rm CD}}$ by counting the zero-photon events in the table. 
Therefore, superconducting nanowire single-photon detectors, which distinguish $\ketbra{0}$ and $\boldsymbol{I}-\ketbra{0}$ can be directly used for two-photon swapping.

When time-bin states are imperfect, the heralded qubit state following photon detection can be obtained via the standard POVM theory. Similarly to Eq.~\eqref{eq:rho-out-time-bin}, the output state can be written as
\begin{align}
    \rho^{\rm out} = &\frac{1}{2}e^{-(\eta/2+N)(|\xi_1|^2+|\xi_2|^2)} \nonumber\\
    &\begin{pmatrix}
        L_n(\eta|\xi_1|^2) &(-\eta \xi_1^*\xi_2)^n/n!\\
        (-\eta \xi_1\xi_2^*)^n/n! & L_n(\eta|\xi_2|^2)
    \end{pmatrix}.\label{eq:rho_out}
\end{align}
From Eq.~(\ref{eq:rho_out}), the product state can be expressed as a $4\times 4$ density matrix $\rho=\rho^{\rm out}_{A A^\prime}\otimes \rho^{\rm out}_{B B^\prime}$.
The measurement POVM consists of photon number states transformed by a $50/50$ beam splitter. For single photon swapping, let 
\begin{align*}
    \hat{M}=U_{\rm BS}^\dagger\ketbra{1010}U_{\rm BS}
\end{align*}
where $\ket{1010}_{A_1B_1A_2B_2}$ represents the photon number state at ports $A$ and $B$ at time bins $1$ and $2$. This measurement corresponds to one of the outcomes that herald state $\ket{\Phi^+_{\rm AB}}$ in Table \ref{table:single-swap}. Since
\begin{align*}
    \hat{\rho} \rightarrow \ketbra{1010} U_{\rm BS} \hat{\rho} U_{\rm BS}^\dagger \ketbra{1010},
\end{align*}
 the post-selected state is given by
 \begin{align*}
     \Tr_{AB}(\hat{M}\hat{\rho}) \ketbra{1010}{1010},
 \end{align*}
up to a normalized constant. Let $\chi_{\hat{M}}(\bxi)$ be the characteristic function of $\hat{M}$, and $\bxi=(\xi_{A_1},\xi_{B_1},\xi_{A_2},\xi_{B_2})$. Combining Eq.~\eqref{eq:BS-transform} and Eq.~\eqref{eq:characterstic-partial-trace}, the explicit expression is given by
\begin{align}
    \chi_{\hat{M}}(\bxi)=&L_1(1/2|\xi_{A_1}-\xi_{B_1}|^2)L_1(1/2|\xi_{A_2}-\xi_{B_2}|^2) \nonumber \\ &e^{-(|\xi_{A_1}|^2+|\xi_{A_2}|^2+|\xi_{B_1}|^2+|\xi_{B_2}|^2)/2}.
    \label{eq:chi_M}
\end{align}
The density matrix of entangled qubits is obtained through element-by-element integration from Eq.~\eqref{eq:rho-AB}:
 \begin{align}
     \rho_{A^\prime B^\prime}\propto \int \diff^8{\bxi} \;\; \chi_{\hat{M}}(-\bxi) (\rho^{\rm out}_{A^\prime A}\otimes \rho^{\rm out}_{B^\prime B}).
     \label{eq:rho-AB1}
 \end{align}
Since the density matrix of $\ketbra{\Phi^+_{A^\prime B^\prime}}$ is
\begin{align}
    \begin{pmatrix}
        0&0&0&0\\
        0&1/2&1/2&0\\
        0&1/2&1/2&0\\
        0&0&0&0
    \end{pmatrix}.
\end{align}
The quantum fidelity \begin{align}
    F&=\bra{\Phi^+_{A^\prime B^\prime}}\hat{\rho}_{A^\prime B^\prime} \ket{\Phi^+_{A^\prime B^\prime}}\nonumber\\
    &=1/2([\rho_{A^\prime B^\prime}]_{22}+[\rho_{A^\prime B^\prime}]_{33}+[\rho_{A^\prime B^\prime}]_{23}+[\rho_{A^\prime B^\prime}]_{32}).
    \label{eq:apd-fidelity}
\end{align}
Substituting Eqs. \eqref{eq:rho_out}, \eqref{eq:chi_M}, and \eqref{eq:rho-AB1}, along with $n=1$ into Eq.~\eqref{eq:apd-fidelity}, we obtain
\begin{subequations}   
\begin{align}
& \rm{Tr}_{A^\prime B^\prime AB}(\hat{M}\hat{\rho})=\frac{(t-1)(t - \eta) [t^2 + 2\eta - t(1 + \eta)]}{2 t^8} 
\nonumber\\
& \qquad \qquad \qquad+ \frac{1}{2} \left[ \frac{3\eta + 2t(t - 1 - \eta)}{2t^4} \right]^2,\\
&{[\rm{Tr}_{A^\prime B^\prime AB}(\hat{M}\hat{\rho})]} F=\frac{[2t(t-1-\eta)+3\eta]^2+\eta^2}{16 t^8},
\end{align}
\label{eq:fidelity-single-photon}
\end{subequations}
where $t=(\eta+1)/2+N$.
When $n=2$ in Eq.~\eqref{eq:rho_out} and $\hat{M}=U_{\rm BS}^\dagger\ketbra{2020}{2020}U_{\rm BS}$, we obtain
\begin{widetext}
\begin{subequations}\label{eq:fidelity-two-photon}
    \begin{align}
        &\rm{Tr}_{A^\prime B^\prime AB}(\hat{M}\hat{\rho})=\frac{1}{32 t^{12}} \left\{ 32 (-1 + t)^4 t^4 - 128 (-2 + t) (-1 + t)^3 t^3 \eta + 16 (-1 + t)^2 t^2 \left[ 43 + 12 (-4 + t) t \right] \eta^2 \right. \nonumber \\
        & \qquad \qquad  \left. - 16 (-1 + t) t \left[ -47 + 2 t \left( 43 + 4 (-6 + t) t \right) \right] \eta^3 + \left[ 289 + 16 t \left( -47 + t \left( 43 + 2 (-8 + t) t \right) \right) \right] \eta^4 \right\},\\
        & {[\rm{Tr}_{A^\prime B^\prime AB}(\hat{M}\hat{\rho})]} F=\frac{1}{32 t^{12}} \left\{ 8 (-1 + t)^4 t^4 - 32 (-2 + t) (-1 + t)^3 t^3 \eta + 12 (-1 + t)^2 t^2 (-5 + 2 t) (-3 + 2 t) \eta^2 \right.\nonumber\\
        &\left. \qquad \qquad - 8 (-2 + t) (-1 + t) t \left[ 13 + 4 (-4 + t) t \right] \eta^3 + \left[ 85 + 4 (-4 + t) t \left( 13 + 2 (-4 + t) t \right) \right] \eta^4 \right\}.        
    \end{align}
\end{subequations}
\end{widetext}

\section{Fidelity by multiple time bins}\label{apd:fidelity-multiple-time-bin}
Let $t=(\eta+1)/2+N$. The normalization constant $K_0$ in Eq.~\eqref{eq:K0} depends on whether $k$ is even or odd. For odd $k=2l+1$ and even $k$, they are given by
\begin{widetext}
\begin{subequations}
    \begin{align}
        K_0&=\frac{1}{4} \left\{ \frac{(t - 1)(t - \eta)[t^2 + 2\eta - t(1 + \eta)]}{t^8} \right\}^l \left[ \frac{2 t^3 - 2 \eta^2 - 2 t^2 (1 + \eta) + t \eta (3 + \eta)}{t^5} \right] + \frac{1}{2} \left[ \frac{3 \eta + 2 t (t - \eta - 1)}{2 t^4} \right]^k,\\
        K_0&=\frac{1}{2} \left\{ \frac{(t-1)(t - \eta) \left[ t^2 + 2\eta - t (1 + \eta) \right]}{t^8} \right\}^{k/2}+ \frac{1}{2} \left[ \frac{3 \eta + 2 t (t - \eta - 1)}{2 t^4} \right]^{k}.
    \end{align}
\end{subequations}
\end{widetext}
In both cases the quantum fidelity is then simply obtained from Eq.~\eqref{eq:K0F} where
\begin{align}
    K_0 F &=\frac{1}{4} \left[ \frac{3 \eta + 2 t (t - \eta - 1)}{2 t^4} \right]^{k} + \frac{1}{4} \left( \frac{\eta}{2 t^4} \right)^{k}.
\end{align}

\end{document}